\documentclass[preprint, aps, pra, showpacs, floatfix, superscriptaddress]{revtex4-1}
\usepackage{graphicx}
\usepackage{amsmath,amsfonts,amssymb,verbatim,float}
\usepackage[usenames]{color}    %for colours
\usepackage{colortbl}
\usepackage{hyperref} % for links in the bibliography
\usepackage{subfigure}
\usepackage{indentfirst}
\usepackage{physics}

\usepackage[normalem]{ulem}

\usepackage{siunitx} % for formatting tables

\newcommand{\btheta}{\boldsymbol{\theta}}
\newcommand{\balpha}{\boldsymbol{\alpha}}
\newcommand{\bnabla}{\boldsymbol{\nabla}}
\newcommand{\bp}{\boldsymbol{p}}
\newcommand{\br}{\boldsymbol{r}}
\newcommand{\bq}{\boldsymbol{q}}
\newcommand{\hbr}{\hat{\boldsymbol{r}}}
\newcommand{\ha}{\hat{a}}
\begin{document}
\thispagestyle{empty}
\title{
%Calculation of ground-state energies of superheavy elements via quantum algorithms
Calculation of the moscovium ground-state energy by quantum algorithms
}
\author{V.~A.~Zaytsev}
\affiliation{
Department of Physics, St. Petersburg State University,
Universitetskaya naberezhnaya 7/9, 199034 St. Petersburg, Russia
}
\author{M.~E.~Groshev}
\affiliation{
Department of Physics, St. Petersburg State University,
Universitetskaya naberezhnaya 7/9, 199034 St. Petersburg, Russia
}
\author{I.~A.~Maltsev}
\affiliation{
Department of Physics, St. Petersburg State University,
Universitetskaya naberezhnaya 7/9, 199034 St. Petersburg, Russia
}
\author{A.~V.~Durova}
\affiliation{
Department of Physics, St. Petersburg State University,
Universitetskaya naberezhnaya 7/9, 199034 St. Petersburg, Russia
}
\author{V.~M.~Shabaev}
\affiliation{
Department of Physics, St. Petersburg State University,
Universitetskaya naberezhnaya 7/9, 199034 St. Petersburg, Russia
}
\affiliation{
B.P. Konstantinov Petersburg Nuclear Physics Institute of National Research Centre ``Kurchatov Institute'', Gatchina, 188300 Leningrad District, Russia
}
%
%\date{18 February 2022}
%
\begin{abstract}
We investigate the possibility to calculate the ground-state energy of the atomic systems on a quantum computer. For this purpose we evaluate the lowest binding energy of the moscovium atom with the use of the iterative phase estimation and variational quantum eigensolver. The calculations by the variational quantum eigensolver are performed with a disentangled unitary coupled cluster ansatz and with various types of hardware-efficient ansatze. The optimization is performed with the use of the Adam and Quantum Natural Gradients procedures. The scalability of the ansatze and optimizers is tested by increasing the size of the basis set and the number of active electrons. The number of gates required for the iterative phase estimation and variational quantum eigensolver is also estimated.
\end{abstract}
\maketitle
%
% ===================================================================
\section{Introduction}
The knowledge about an electron structure is crucial for understanding the properties of atoms, chemical compounds, and materials. 
Despite the impressive progress in calculation methods and computer hardware during the last few decades, the calculation of electron correlations is still a challenging problem even for the atomic systems. 
Meanwhile, the accurate description of atomic electron structure would have a tremendous impact on the fundamental physics and the foundation of chemistry. 
For instance, the tests of the Standard Model in atomic systems are now limited by the electron-correlation contribution~\cite{Shabaev_2005, Dzuba_2012}. 
In order to predict the chemical properties of superheavy elements with many valence electrons, one also has to accurately model systems with large numbers of interacting electrons that is rather problematic~\cite{Kaygorodov_2021, Kaygorodov_2022}. 
\\ 
\indent
There is a fundamental limitation which prevents an accurate description of many-electron systems governed by quantum mechanics on the classical computers.
The limitation is caused by the complexity of the system wave function which grows exponentially with the number of particles. 
As a result, {\it ab initio} calculations of many-electron systems is often impossible and the approximate methods, such as, e.g., density functional theory, are used. 
The accuracy of the approximate methods, however, is quite limited.
\\ 
\indent
As it was firstly proposed by Manin~\cite{Manin_80} and Feynman~\cite{Feynman_82}, a natural way to overcome the problem of exponentially growing complexity is to calculate the quantum systems on the quantum devices. 
In order to perform such calculations, one needs completely new algorithms which can exploit the power of quantum computers. 
To date several quantum algorithms for electronic structure calculations have been developed~(see reviews~\cite{McArdle_2020, Bauer_CR120_12685_2020} and references therein). 
The first method, proposed in Ref.~\cite{Abrams_99}, was based on the quantum phase estimation algorithm (PEA)~\cite{Kitaev} combined with the Trotterization of the evolution operator~\cite{Trotter, Suzuki}. 
The PEA requires additional qubits (ancilla), whose number is defined by the desired accuracy.
It is possible, however, to implement PEA in an iterative manner (iPEA) with reduction of the ancilla qubit number to one~\cite{Griffiths_PRL76_3228_1996, Parker_PRL85_3049_2000}. 
If there were an ideal quantum computer, the PEA would provide an exponential advantage over its classical counterparts. 
However, the modern quantum hardware suffers from the decoherence and other types of noise and cannot execute long sequences (circuits) of operations (gates) with a correct output.
Meanwhile, the PEA requires very deep circuits and, therefore, it is unlikely to demonstrate the advantage over the classical algorithms in the nearest future. 
\\
\indent
A more promising algorithm for near-term noisy quantum computers is the variational quantum eigensolver (VQE) which was firstly proposed in Ref.~\cite{Peruzzo_NC5_4213_2014}.
The VQE has the hybrid quantum-classical nature and is based on minimization of a Hamiltonian expectation value via tuning the trial wave function with some parameterized unitary operator which is called an ``ansatz''.
The quantum computer is used for the state preparation according to the ansatz and the measurement of expectation values. The optimization procedure is performed on the classical computer. 
The VQE requires much shallower circuits than the PEA and, moreover, it is more robust to the noise due to the variational nature~\cite{Malley_16, Barkoutsos_18}. 
The noise resilience of the VQE made it possible to calculate on the real quantum devices the energies of simple molecules~(see, for instance, Refs.~\cite{Peruzzo_NC5_4213_2014, Malley_16, Kandala_N549_242_2017, Shen_17, Hempel_18, Ollitrault_20, Nam_20, Tilly_2020}),
molecular dynamics~\cite{Sokolov_Barkoutsos_2021}, and deuteron binding energy~\cite{Dumitrescu_2018}.
All these calculations were carried out in very small basis sets and the applicability of the VQE for the large-scale calculations is still an open question. 
The ansatz and the optimization algorithm are both crucial for the scalability of the VQE and its potential to eventually demonstrate the advantage over the classical methods.
%
%\\ \indent
% The first problem which should be solved for successful realization of the VQE is the appropriate choice of the ansatz. 
% To date different strategies for {\rd the} ansatz construction were proposed. 
% Among them one can distinguish the hardware-efficient and the problem inspired ansatze. 
% The first group is designed to be easily implemented with the hardware. 
% The later one takes into account the intrinsic features of the modeled system. 
% The most often considered problem inspired ansatz is the unitary coupled cluster (UCC) ansatz, which can be viewed as an extension of the widely used Coupled Cluster method. 
% Another challenge is to find the global minimum with the classical optimization procedure.
% It is not clear that the required optimization is possible for the large systems.
% The ansatz and the optimization algorithm are both crucial for the scalability of the VQE and its potential to eventually demonstrate the advantage over the classical methods.
%
\\ \indent
The aim of the present work is to investigate the possibility to calculate the electronic structure of the atomic systems by the quantum algorithms. 
We focused on the calculation of the ground-state energy of the moscovium atom. 
Moscovium is chosen due to the half-filled $p$-shell in the ground configuration.
This ground configuration makes it difficult to calculate the ground-state energy with the use of the standard Coupled Cluster method, often referred to as the ``golden standard'' of quantum chemistry.
%We employ the VQE algorithm since it is the most promising for the near-term quantum devices.
% The VQE was evaluated in combination with two different types of hardware-efficient ansatze and with the problem inspired UCC ansatz alongside with different optimization procedures.
Here we apply the VQE with the problem-inspired disentangled unitary coupled cluster ansatz~\cite{Evangelista_JCP151_244112_2019} and with different types of hardware-efficient ansatze~\cite{Kandala_N549_242_2017}.
The optimization is performed with the use of the Adam~\cite{Kingma_Ba} and Quantum Natural Gradients~\cite{Stokes_Q4_269, Wierichs_PRR2_043246_2020} algorithm.
We also apply iPEA, which, in contrast to the VQE, is free of the ansatz selection problems and optimization difficulties and, moreover, will provide the exact value for the ground-state energy if launched on an ideal quantum computer.
The number of gates required by the VQE algorithm and iPEA to achieve the same level of accuracy is estimated.
%
% ===================================================================
%
\section{BASIC FORMALISM}
In the second quantization, the electronic structure Hamiltonian is given by
\begin{equation}
\hat{H} = 
\sum_{p,q} h_{pq} \ha^\dagger_p \ha_q
+
\frac{1}{2}
\sum_{p,q,r,s} h_{pqrs} \ha^\dagger_p \ha^\dagger_q \ha_r \ha_s,
\label{eq_H}
\end{equation}
where the annihilation $\ha_p$ and creation $\ha_p^\dagger$ operators correspond, respectively, to removing and adding an electron described by the relativistic one-electron orbital $\psi_p$.
Here we assume that these orbitals are defined by the principal quantum number $n$, parity $l$, total angular momentum $j$ and its projection $m$.
The one-electron integrals are given by
\begin{equation}
h_{pq} = \int d\br \psi_p^\dagger(\br) \hat{h}_{D}(r) \psi_q(\br),
\end{equation}
where
\begin{equation}
\hat{h}_{D} = c\balpha\cdot\hat{\bp} + (\beta - 1)m_ec^2 + V_{\rm nuc}(r)
\end{equation}
is the one-electron Dirac Hamiltonian
with $\balpha$ and $\beta$ being the Dirac matrices, $\hat{\bp}$ is the momentum operator, and $V_{\rm nuc}$ is the nuclear potential.
The two-electron integrals are defined as
\begin{equation}
h_{pqrs} = \int d\br_1 d\br_2
\psi_p^\dagger(\br_1) \psi_q^\dagger(\br_2) 
\left[ V_{C}(r_{12}) + V_{B}(r_{12}) \right]
\psi_s(\br_1) \psi_r(\br_2)
\end{equation}
with the Coulomb and Breit interelectronic-interaction operators given by 
\begin{equation}
V_{C}(r_{12}) = \frac{\alpha c}{r_{12}},
\end{equation}
\begin{equation}
V_{B}(r_{12}) =
-\frac{\alpha c}{2r_{12}} 
\left[\balpha_1\cdot\balpha_2
+ 
\left(\balpha_1\cdot\hbr_{12}\right)
\left(\balpha_2\cdot\hbr_{12}\right)
\right],
\end{equation}
respectively. 
Here $\hbr_{12} = \br_{12} / r_{12}$ with $\br_{12} = \br_1 - \br_2$ and $r_{12} = |\br_{12}|$.
The one- and two-electron integrals are calculated on a classical computer and serve 
as input parameters to the quantum simulation.
\\
\indent
The ground-state wave function of a many-electron system described by the Hamiltonian~\eqref{eq_H} can be expressed as a linear superposition of the Slater determinants $\ket{\Phi_n}$,
\begin{equation}
\ket{\Psi_0} = \sum_n c_n \ket{\Phi_n}.
\end{equation}
In the second quantization formalism, the Slater determinant is given by
\begin{equation}
\ket{\Phi_n}
= \ket{f_0 \dots f_{N_{\rm orbs}-1}}
= \prod_{i = 0}^{N_{\rm orbs}-1} \left(a_i^\dagger\right)^{f_i} \ket{vac},
\end{equation}
where $N_{\rm orbs}$ is the number of one-electron orbitals in the basis set and $f_i \in \lbrace0,1\rbrace$ stands for the occupation number of the orbital $i$.
We note that the ground-state wave function and the Slater determinants are eigenstates of the particle number, total angular-momentum, and parity operators.
\\
\indent
To simulate the system described by the Hamiltonian~\eqref{eq_H} on a quantum computer, one needs to build a correspondence between the orbitals-based Fock space and the qubit space
\begin{equation}
\vert f_{0} \dots f_{N_{\rm orbs}-1} \rangle  
\rightarrow 
\ket{\bq}
\equiv
\vert q_{N_{\rm orbs}-1} \rangle \otimes \dots \otimes \vert q_0 \rangle
=
\ket{ q_{N_{\rm orbs}-1}, \dots , q_0 }.
\end{equation}
The most straightforward way is to use the Jordan-Wigner mapping~\cite{Jordan_Wigner}, which associates qubit state $\ket{q_i}$ with the occupation number $f_i$, i.e., the qubit states $\vert0\rangle$ and $\vert1\rangle$ correspond to the empty and filled state, respectively.
Here we utilize the parity encoding~\cite{Bravyi_Kitaev}, which is more appropriate for the simulation of the electronic structure of the atom.
In this mapping, the qubit state $\ket{q_j}$ stores the parity of the number of filled orbitals up to orbital $j$, that is
\begin{equation}
p_j = \left(\sum_{i=0}^j f_i\right) {\rm mod}\ 2.
\end{equation}
The annihilation and creation operators, obeying fermionic anti-commutation relations, in terms of the qubit operations are given by 
\begin{eqnarray}
\hat{a}_j & = & 
\sigma^x_{N_{\rm orbs}-1} \otimes \dots \otimes \sigma^x_{j+1} \otimes \mathcal{P}^-_j,
\label{eq_ann}
\\
\hat{a}_j^\dagger & = & 
\sigma^x_{N_{\rm orbs}-1} \otimes \dots \otimes \sigma^x_{j+1} \otimes \mathcal{P}^+_j,
\label{eq_crt}
\end{eqnarray}
where
\begin{equation}
\mathcal{P}^\pm_j 
= 
\frac{1}{2} \left( \sigma^x_j\otimes \sigma^z_{j-1} \mp i\sigma^y_j \right)
\end{equation}
and
\begin{equation}
\sigma^x = \begin{pmatrix}0 && 1 \\ 1 && 0\end{pmatrix},
\quad
\sigma^y = \begin{pmatrix}0 && -i \\ i && 0\end{pmatrix},
\quad
\sigma^z = \begin{pmatrix}1 && 0 \\ 0 && -1\end{pmatrix}
\end{equation}
are the Pauli matrices.
Here it is assumed that the identity operator $I$ is applied to the qubits, for which the operations are not explicitly indicated.
Substituting Eqs.~\eqref{eq_ann} and \eqref{eq_crt} into Eq.~\eqref{eq_H}, one obtains the Hamiltonian in the form 
\begin{equation}
\hat{H} = \sum_{n} \gamma_n P_n,
\label{eq_ham_ps}
\end{equation}
where $P_n \in \lbrace I, \sigma^x, \sigma^y, \sigma^z \rbrace^{\otimes N_{\rm orbs}}$ is the tensor product of Pauli matrices, so-called Pauli string (PS).
\\
\indent
In the parity mapping the ordering of the orbitals is important.
Here we place odd orbitals before even ones.
In such ordering, the parity of the whole system $P$ is stored in the qubit corresponding to the last odd orbital in the basis.
If $P$ is known, which is most often the case in atomic calculations, it can be fixed by tapering off this qubit~\cite{Bravyi}. 
We also taper the last qubit, thus fixing the parity of the particle number.
\\
\indent
%
%Having constructed the correspondence between the qubits and the electronic structure of the atomic system described by the Hamiltonian~\eqref{eq_H}, we now turn to the calculation of the ground-state energy.
%
%Here we perform the calculation by the variational quantum eigensolver (VQE) and iterative phase estimation algorithm (iPEA).
Having constructed the correspondence between the qubits and the electronic structure 
of the atomic system, we now turn to the quantum algorithms. In the present work, 
we consider the variational quantum eigensolver (VQE) and iterative phase estimation algorithm (iPEA)
for calculation of the ground-state energy.

% -------------------------------------------------------------------
%
\subsection{Variational Quantum Eigensolver (VQE)}
%
% \begin{itemize}
% \item VQE is the iterative hybrid quantum-classical algorithm 
% \item On each iteration, first, the quantum computer is used to prepared the trial state which has to approximate the ground state. 
% \item Next, the expectation value is measured on a quantum computer
% \item new parameters theta which are expected to reduce are determined on 
% \end{itemize}
% The variational quantum eigensolver (VQE), which was first proposed in Ref.~\cite{Peruzzo_NC5_4213_2014}, is the iterative hybrid quantum-classical algorithm.
% Each iteration starts from the preparation of the trial state 
% %
% \begin{equation}
% \ket{\Psi\left(\btheta\right)}
% = U\left(\btheta\right) 
% \ket{\Psi_0}
% \label{eq_trial}
% \end{equation}
% %
% on a quantum computer. 
% The explicit form of the parameterized unitary operator $U\left(\btheta\right)$, which is called an ``ansatz'', and of the initial state $\ket{\Psi_0}$ will be specified below.
% parametrized guess
% intended to 
% approximate 
% close to the solution
% good at describing 
% / the ground-state wave function / the solution of interest
% ----------------------------
%
The VQE is the hybrid quantum-classical algorithm. 
It was first proposed in Ref.~\cite{Peruzzo_NC5_4213_2014} and constitutes in minimization of the expectation value
\begin{equation}
E_{\btheta} 
=
\frac{ \expval{\hat{H}}{\Psi\left(\btheta\right)} }
{ \braket{\Psi\left(\btheta\right)} }
\label{eq_expval},
\end{equation}
with the trial wave function defined as
\begin{equation}
\ket{\Psi\left(\btheta\right)}
= U\left(\btheta\right) 
\ket{\Psi'}.
\label{eq_trial}
\end{equation}
The explicit form of the parameterized unitary operator $U\left(\btheta\right)$, which is called an ``ansatz'', and of the initial state $\ket{\Psi'}$ will be specified below.
The VQE algorithm proceeds as follows.
First the parameterized guess of the ground-state wave function~\eqref{eq_trial} is prepared and the expectation value~\eqref{eq_expval} is measured on the quantum computer.
On the next step, the parameters $\btheta$ are updated by a classical optimization algorithm which takes on input $E_{\btheta}$.
The adjusted parameters are further used by the quantum computer in the new iteration.
The procedure is repeated until convergence is reached.
\\
\indent
The ansatz $U(\btheta)$ is one of the most important components of the VQE algorithm.
First, the wave function $\ket{\Psi(\btheta)}$ constructed by the ansatz should provide a good approximation to the ground-state wave function.
% One of the quality criteria is the ability to obtain a good approximation to the ground-state wave function.
The number of the parameters required for the construction, meanwhile, should grow polynomially with the growth of the system and basis sizes.
Moreover, the ansatz has to consist of unitaries which can be readily transformed into the operations allowed on the quantum computer.
The literature comprises various strategies for the ansatz construction.
For example, the ansatz structure may be fixed from the beginning, or it can be changed adaptively in course of the calculation.
One can also classify ansatze as problem inspired or hardware efficient.
More details on these and many other strategies can be found in Refs.~\cite{Sokolov_Barkoutsos_2020, Takeshita_Rubin_2020, Zhang_Gomes_2021} and in the reviews~\cite{Bauer_CR120_12685_2020, Cerezo_NRP3_625_2021, Bharti_RMP94_015004_2022}.
%
%\\
%\indent
%
Here we use disentangled unitary coupled cluster~\cite{Evangelista_JCP151_244112_2019} and hardware efficient~\cite{Kandala_N549_242_2017} ansatze, which are problem oriented and device specific, respectively.
In both cases the structure of the ansatze remains fixed throughout the calculation.
%
% - - - - - - - - - - - - - - - - - - - - - - - - - - - - - - - - - -
%
\subsubsection{Disentangled Unitary Coupled Cluster}
The Unitary Coupled Cluster (UCC) approach obeys the variational principle and, thus, can be applied to a much broader range of systems when compared with the conventional Coupled Cluster method~\cite{Taube}.
Unfortunatelly, the UCC parametrization is exponentially costly to implement without approximation on a classical computer. 
This approach, hovewer, can be effectively implemented on a quantum computer~\cite{Romero}.
In the UCC approach, the generic state is expressed as
\begin{equation}
\ket{\Psi_{\rm UCC}} = 
e^{\hat{T} - \hat{T}^\dagger}
\ket{\Psi'}.
\label{eq_UCC}
\end{equation}
Here
\begin{equation}
\hat{T} 
=
\sum_{\mu}^{\rm exc}t_\mu \hat{\tau}_\mu
\label{eq_exc_T}
\end{equation}
is the excitation operator that moves the electrons from the occupied orbitals of the reference Slater determinant $\ket{\Psi'}$ to unoccupied ones.
In Eq.~\eqref{eq_exc_T}, the summation is performed over all unique excitations, $t_\mu$ is the cluster amplitude, and 
\begin{equation}
\hat{\tau}_\mu 
\equiv \ha_{ij\dots}^{ab\dots}
= \ha_a^\dagger \ha_b^\dagger \dots \ha_j \ha_i,
\label{eq_tau}
\end{equation}
where $\mu$ denotes indices $i,j,\dots$ and $a,b,\dots$, which label occupied and virtual orbitals of the reference state, respectively.
Here and throughout we assume that the wave function and cluster amplitudes are real, that allows us to write the power of the exponent in Eq.~\eqref{eq_UCC} as follows
\begin{equation}
\hat{T} - \hat{T}^\dagger
=
\sum_{\mu}^{\rm exc}t_\mu \left(
\hat{\tau}_\mu - \hat{\tau}_\mu^\dagger 
\right).
\label{eq_TT}
\end{equation}
\indent
Substituting Eqs.~\eqref{eq_ann} and~\eqref{eq_crt} into Eq.~\eqref{eq_tau}, one can express $\hat{T} - \hat{T}^\dagger$ as a linear superposition of PSs.
At present, unfortunately, there exist no effective methods for the conversion of this ansatz into 
%the qubit operations.
the quantum gates.
Here, to avoid this problem, we use an alternative formulation of the UCC method with the factorized exponent
\begin{equation}
\ket{\Psi_{\rm dUCC}} 
= 
\prod_i e^{t_{\mu_i} (\tau_{\mu_i} - \tau_{\mu_i}^\dagger)} \ket{\Psi'}.
\end{equation}
In this disentangled UCC (dUCC) ansatz, which was proposed in Ref.~\cite{Evangelista_JCP151_244112_2019}, each exponential term appears exactly once and the special ordering of these terms is used.
We note that the PSs corresponding to a given difference $\tau_\mu - \tau_\mu^\dagger$ commute with each other, and, thus, can be easily converted into the gates.
For example, one can use the ``greedy'' algorithm 
%described in many textbooks 
(see, e.g., Ref.~\cite{Nielsen_Chuang}).
Here, instead, we follow the three-step strategy proposed in Refs.~\cite{Cowtan, Vandaele, Patel} which allows us to substantially reduce the required number of gates compared to the ``greedy'' algorithm.
The details of this strategy and ``greedy'' algorithm are briefly summarized in Appendix~\ref{appendix}.
\\
\indent
Though the dUCC ansatz can be easily converted into the quantum gates, 
the number of the cluster amplitudes grows exponentially with the basis size if all unique excitations are included. 
This growth results in a proportional increase of the required computational (both classical and quantum) resources, that makes the exact calculations unfeasible.
In the present work, we restrict ourselves to the single and double excitations and refer to such ansatz as dUCC-SD.
This ansatz is equivalent to the conventional UCC-SD ansatz being factorized with the use of the low-order Trotter approximation~\cite{Trotter, Suzuki}.
The approximate factorization is not uniquely defined and, as was shown in Ref.~\cite{Grimsley_JCTC16_1_2020}, the ordering of the operators affects the accuracy that can be achieved for the energy.
In the present work, we arrange the single and double excitation operators in the order as they appear in the dUCC ansatz~\cite{Evangelista_JCP151_244112_2019}.
% inefficient to represent on a classical computer [Yung, M.-H., J. Casanova, A. Mezzacapo, J. McClean, L. Lamata, A. Aspuru-Guzik, and E. Solano, Sci. Rep. 4, 3589 (2014)]
%
% - - - - - - - - - - - - - - - - - - - - - - - - - - - - - - - - - -
%
\subsubsection{Hardware Efficient}
\begin{figure}[h!]
\begin{minipage}[b]{0.45\textwidth}
\centering
\includegraphics[width=\textwidth]{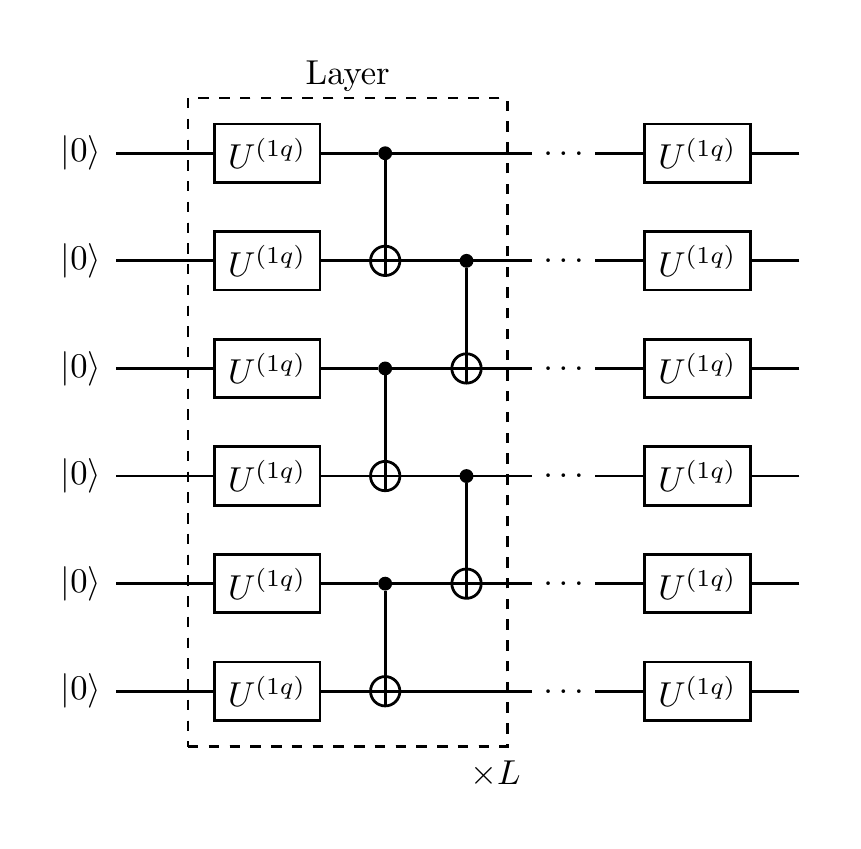}
\end{minipage}
\hfill
\begin{minipage}[b]{0.53\textwidth}
\centering
\includegraphics[width=\textwidth]{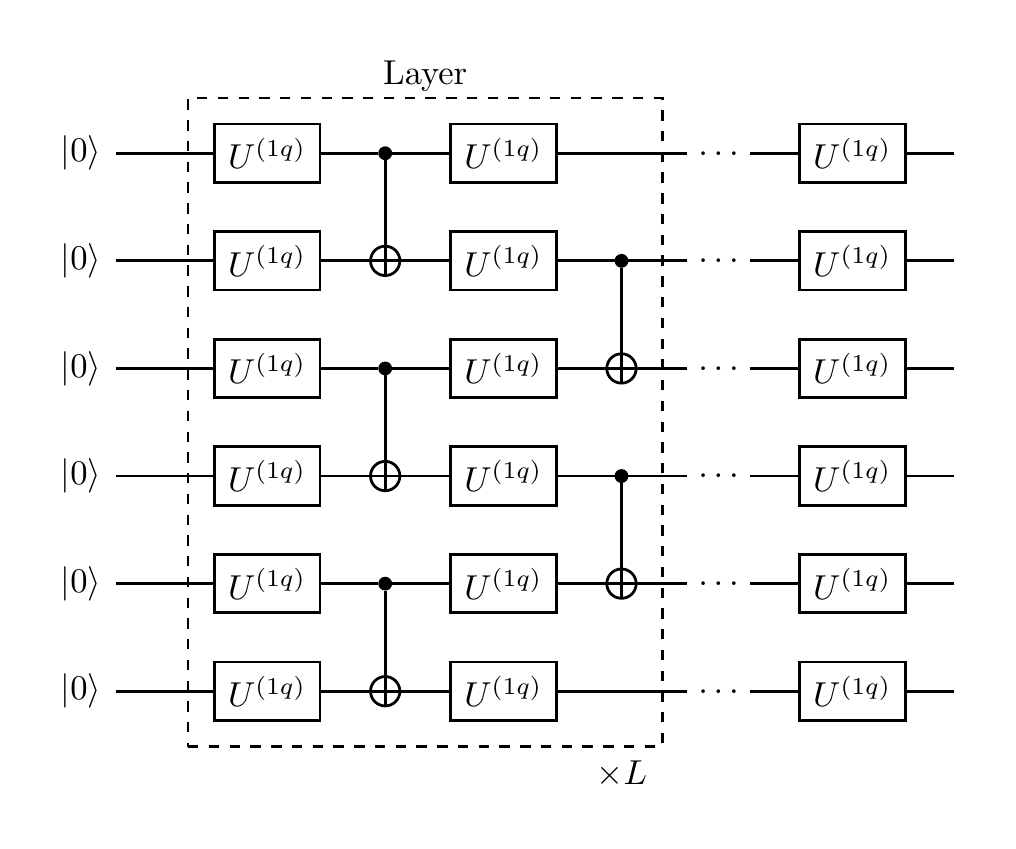}
\end{minipage}
\caption{
Schematic representation of the hardware effecient ansatze with $L$ layers. 
Left and right panels are, respectively, correspond to the ansatze with two different entangling layers, merged and splitted by a layer of one-qubit rotations. 
}
\label{fig_he_ansatz}
\end{figure}

The problem-oriented ansatz, such as dUCC, typically requires a large number of gates and all-to-all connectivity of the qubits.
Unfortunately, currently it is impossible to study the systems intractable for classical approaches on quantum computers with such ansatze.
The calculations on the modern noisy quantum devices~\cite{Preskill} 
are strongly limited by the restricted gate set, specific qubit connectivity, operations fidelities, coherence time, and other imperfections.
To overcome these constraints the so-called hardware-efficient (HE) ansatz was proposed in Ref.~\cite{Kandala_N549_242_2017}.
This ansatz has the form
\begin{equation}
U_{\rm HE}(\btheta) 
= 
\prod_{q = 1}^{N_q} U^{(1q)}_q(\btheta_{L+1q})
\times
U_{\rm ent} \prod_{q = 1}^{N_q} U^{(1q)}_q(\btheta_{Lq})
\times
\dots
\times
\underbrace{
U_{\rm ent} \prod_{q = 1}^{N_q} U^{(1q)}_q(\btheta_{1q})
}_{\rm layer},
\end{equation}
where $N_q$ is the number of qubits. Each layer of the ansatz 
consists of the non-parameterized enatangling operator $U_{\rm ent}$ and of the single-qubit rotation gates $U^{(1q)}_q(\btheta_{lq})$.
The unitary $U_{\rm ent}$ is constructed from architecture specific entangling gates applied to directly connected qubits.
The hardware efficient ansatz is not related to the problem to be solved and, thus, the initial wave function can be of arbitrary form and is usually chosen as 
\begin{equation}
\ket{\Psi'} = \ket{\bf 0} \equiv \ket{0}^{\otimes N_q}.
\end{equation}
\\
\indent
In the present work, we consider hardware efficient ansatze, which differ by the structure of the layers and by the single-qubit rotation gates.
Specifically, the ansatze with two different entangling layers being (i) merged and (ii) splitted by a layer of one-qubit rotations (see Fig.~\ref{fig_he_ansatz}) are used.
As single-qubit operations we utilize
\begin{equation}
U^{(1q)}_{zyz} = R_z(\theta_3)R_y(\theta_2)R_z(\theta_1)
\end{equation}
and
\begin{equation}
U^{(1q)}_{zx} = R_z(\theta_2)R_x(\theta_1),
\end{equation}
where
\begin{equation}
R_x(\theta) = e^{-i\sigma^x\theta/2} 
= 
\begin{pmatrix}
\cos\frac{\theta}{2} & -i\sin\frac{\theta}{2}
\\
-i\sin\frac{\theta}{2} & \cos\frac{\theta}{2}
\end{pmatrix},
\end{equation}
\begin{equation}
R_y(\theta) = e^{-i\sigma^y\theta/2} 
= 
\begin{pmatrix}
\cos\frac{\theta}{2} & -\sin\frac{\theta}{2}
\\
\sin\frac{\theta}{2} & \cos\frac{\theta}{2}.
\end{pmatrix},
\end{equation}
and
\begin{equation}
R_z\left(\theta\right) 
= e^{-i\sigma^z\theta/2} 
= \begin{pmatrix}
e^{-i\theta/2} & 0 \\ 0 & e^{i\theta/2}
\end{pmatrix}.
\end{equation}
%
% -------------------------------------------------------------------
%
\subsection{The iterative phase estimation algorithm (iPEA)}

\begin{figure}[ht]
\centering
\includegraphics[width=\textwidth]{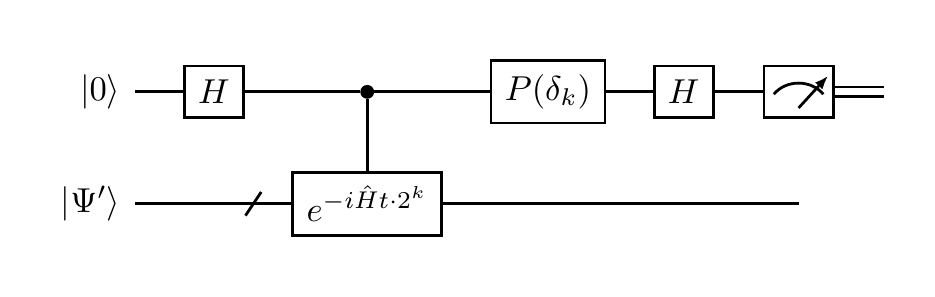}
\caption{
The quantum circuit of the iPEA for obtaining the $k$th bit.
}
\label{fig_ipea_circ}
\end{figure}

In the present work, we use the iterative version of the PEA, which requires one ancilla qubit and involves one controlled evolution operation~\cite{Griffiths_PRL76_3228_1996,Parker_PRL85_3049_2000,Whitfield}. 
In the iPEA, the energy is measured bitwise starting from the least significant bit. 
The quantum circuit for the measurement of the $k$th bit is presented in Fig.~\ref{fig_ipea_circ}.
In the figure,
\begin{equation}
P(\delta) = 
\begin{pmatrix} 1 && 0 \\ 0 && e^{i\delta} \end{pmatrix},
\qquad
H = \frac{1}{\sqrt{2}}\begin{pmatrix}1 & 1 \\1 & -1 \end{pmatrix},
\end{equation}
and $\ket{\Psi'}$ is the initial approximation for the ground state $\ket{\Psi_0}$.
The state $\ket{\Psi'}$ can be decomposed into the exact Hamiltonian eigenstates as
\begin{equation}
\ket{\Psi'} = \sum_{n} c_n \ket{\Psi_n}.
\end{equation}
The circuit shown in Fig.~\ref{fig_ipea_circ} modifies the state of the quantum computer as
\begin{equation}
\ket{0}\ket{\Psi'}
\xrightarrow[]{\rm iPEA}
\frac{1}{2}
\sum_n c_n 
\left[
\left(1 + e^{-iE_nt2^k + i\delta_k}\right) \ket{0}
+
\left(1 - e^{-iE_nt2^k + i\delta_k}\right) \ket{1}
\right]
\ket{\Psi_n}.
\end{equation}
Here $E_n$ are the Hamiltonian eigenvalues, 
the propagation time $t$ and the phase $\delta_k$ are chosen as follows.
First, one needs to fix the interval $\Delta E$ containing the ground state energy and fix one of the endpoints of this interval $E'$.
Then the ground state energy $E_0$ can be represented in the binary form as
\begin{equation}
-\frac{E_0 - E'}{\Delta E} 
= 
0\cdot b_0 b_1 \dots {}
=
\sum_{i=0}^{N_{\rm bits}-1}\frac{b_i}{2^{i + 1}} + \varepsilon,
\label{eq_binary}
\end{equation}
where each $b_i$ is zero or one, $N_{\rm bits}$ is the number of the accounted bits in the expansion, $\varepsilon \leqslant 1/2^{N_{\rm bits}+1}$ designates the truncation error, and it is assumed that $E_0 < E' < 0$.
By choosing 
\begin{equation}
t = \frac{2\pi}{\Delta E},
\label{eq_ipea_t}
\end{equation}
and 
\begin{equation}
\delta_k 
= E't2^k 
- 2\pi\left(\frac{b_{k+1}}{2^2} + \frac{b_{k+2}}{2^3} + \dots + \frac{b_{N_{\rm bits}-1}}{2^{N_{\rm bits} - k}}\right)
= 
E't2^k 
-
2\pi\sum_{i = k+1}^{N_{\rm bits}-1} \frac{b_{i}}{2^{i+1-k}},
\label{eq_phase_ipea}
\end{equation}
one can show that
\begin{equation}
\ket{0}\ket{\Psi'}
\xrightarrow[]{\rm iPEA}
c_0 e^{i\xi_k}
\left[ \cos(\xi_k)\ket{b_{k}} -i \sin(\xi_k)\ket{\bar{b}_{k}} \right] \ket{\Psi_0}
+ 
\sum_{n \neq 0}
\left[ \alpha^{(+)}_{nk} \ket{0} + \alpha^{(-)}_{nk} \ket{1} \right]
\ket{\Psi_n},
\end{equation}
with
\begin{eqnarray}
\xi_k & = & 2^k\varepsilon\pi,
\\
\alpha^{(\pm)}_{nk} & = & \frac{1 \pm e^{-iE_nt2^k + i\delta_k}}{2}.
\end{eqnarray}
Neglecting the error $\varepsilon$ and assuming large overlap of the initial and ground states $c_0$, one can see that the iPEA algorithm allows us to sequentially measure the bits of the decomposition~\eqref{eq_binary}.
Indeed, to measure the last bit, one needs to set $k = N_{\rm bits} - 1$.
The measured bit determines the phase~\eqref{eq_phase_ipea} which is used in the measurement of the next bit with $k = N_{\rm bits} - 2$.
The process is repeated until all bits are measured.
%Note that the presence of the error $\varepsilon$ makes the measured value of the last bit random.
%The influence of the error on the measurement of the next bits decreases rapidly if the least significant bits are chosen correctly.
%
\\
\indent
In order to apply the iPEA, one needs to construct the controlled evolution operator, that can be easily performed if the evolution operator itself is already expressed in terms of the qubit operations. 
At present, however, there exists no effective method for the transformation of the evolution operator into the gates.
This is related to the fact that in the qubit space the Hamiltonian is represented as the linear combination of PSs~\eqref{eq_ham_ps}, which, in the general case, do not commute with each other.
Therefore, the evolution operator cannot be directly converted into the product of exponentials, which can be transformed into 
the gates.
In the present work, we transform the evolution operator with using the first-order Suzuki-Trotter formula~\cite{Trotter, Suzuki}:
\begin{equation}
e^{-i\hat{H}t}
= e^{-i\sum_{n}\gamma_n P_n t}
= \left(e^{-i\sum_{n}\gamma_n P_n \tau}\right)^{N_{t}}
\approx \left(e^{-i\hat{H}'\tau}\right)^{N_{t}} + O\left(t\tau\right),
\label{eq_evolution}
\end{equation}
where $\tau = t / N_t$ and $N_t$ is the so-called Trotter number and $H'$ is defined by
\begin{equation}
e^{-i\hat{H}'\tau} = \prod_n e^{-i \gamma_n P_n \tau}.
\label{eq_evolution_trot}
\end{equation}
The evolution operator in the form~\eqref{eq_evolution} can be easily converted into qubit operations, and, as a result, the construction of the controlled evolution can be directly performed.
We note that the use of the Suzuki-Trotter formula corresponds to the change of the Hamiltonian $\hat{H}$ by $\hat{H}'$.
The accuracy of such approximation increases with the growth of $N_t$.
%
% ===================================================================
%
\section{RESULTS AND DISCUSSION}
In the present work, we study the possibility to calculate the ground-state energy of ionic and atomic systems by the quantum algorithms VQE and iPEA on the example of the moscovium atom (Mc, $Z=115$). 
The choice of Mc is explained by its ground state configuration $6d^{10} 7s^2 7p^3$, which has the half-filled $p$-shell, that makes the system difficult for calculation with the conventional CC approach.
\\ \indent
The one- and two-electron integrals are evaluated in the basis of Dirac-Fock-Sturm orbitals~\cite{Tupitsyn_OS_2003, Tupitsyn_PRA_2003, Tupitsyn_PRA_2005, Tupitsyn_PRA_2018} 
and passed to the quantum algorithms as input parameters.
We consider three different basis sets defined as shown in Table~\ref{tb_basis}.
In the basis sets $A$ and $B$, the shell $6d^{10}$ is frozen and the active space consists of 5 electrons in 8 and 16 orbitals with given angular momentum projection, respectively.
In the largest $C$ basis the shell $6d^{10}$ is unfreezed that corresponds to 15 active electrons in 26 orbitals.
\begin{table}[h]
\begin{ruledtabular}
\caption{
The radial orbitals (second column) used for the construction of the basis set are listed as the increments with respect to the preceding basis set.
For each basis the number of active electrons $N_e$, the number of orbitals with the given angular momentum projection $N_{\rm orbs}$, and the number of the required qubits $N_q$ are given.
}
\begin{tabular}{lllll}
Basis & Radial Orbitals & $N_e$ & $N_{\rm orbs}$ & $N_q$
\\ \hline
$A$ & $\{7s,7p_{1/2},7p_{3/2}\}$ & 5 & 8 & 6
\\
$B$ & $+\{8s, 8p_{1/2}, 8p_{3/2}\}$ & 5 & 16 & 14
\\
$C$ & $+\{6d_{3/2}, 6d_{5/2}\}$ & 15 & 26 & 24
\end{tabular}
\label{tb_basis}
\end{ruledtabular}
\end{table}
Since in the parity encoding two qubits can be tapered off, the number of qubits is $N_q = N_{\rm orbs} - 2$.
Simulation of the chosen quantum algorithms for basis sets with more $N_e$ and $N_{\rm orbs}$ than in the $C$ basis requires too much computational resources.Therefore, basis sets larger than the $C$ basis are not considered here.
%
% -------------------------------------------------------------------
%
\subsection{VQE}
In the VQE algorithm, one needs to minimize the expectation value~\eqref{eq_expval} by selecting the parameters $\btheta$.
Here we use the Adam~\cite{Kingma_Ba} and quantum natural gradients~\cite{Stokes_Q4_269, Wierichs_PRR2_043246_2020} optimization strategies for such multivariate optimization. 
%
%\\ \indent
%
The Adam optimizer is the first-order gradient descent method, in which each parameter $\theta_i$ is updated individually.
% with an individual adaptively varying coefficient of learning. 
That provides the sufficiently fast convergence and, in addition, eliminates the overshooting and oscillation issues.
%occurring at high learning rates. 
%{\rd The terms "coefficient of learning" and  "learning rates" are not explained here, have not been introduced before and are not known to the general reader. Fix, please.}
Another crucial feature of the Adam optimizer is the momentum, which allows to escape from some local minima in analogy, for example, with a ball in a landscape with friction.
These features make the Adam optimizer one of the most widely used optimization strategies for a broad range of applications.
\\ \indent
The quantum natural gradient (QNG) is the second-order optimization strategy that requires the knowledge of the Hessian of the expectation value~\eqref{eq_expval}.
This method is designed in a way to move in the direction of the steepest descent in the landscape determined by the ansatz.
We note that the gradients and the Hessian of the expectation value~\eqref{eq_expval} can be measured on the quantum computer with the use of the parametric shift rules, whose detailed study is widely represented in the literature (see, e.g., Ref.~\cite{Schuld_PRA99_032331_2019, Mitarai_PRR1_013006_2019, Kyriienko_PRA104_052417_2021} and the recent review~\cite{Wierichs_Q6_677_2022}).
The details of both these optimization methods are presented in Appendix~\ref{appendix_adam_qng}.
%
% - - - - - - - - - - - - - - - - - - - - - - - - - - - - - - - - - -
%
\subsubsection{Hardware Efficient ansatz}
\begin{figure}
\centering
\includegraphics[width=\textwidth]{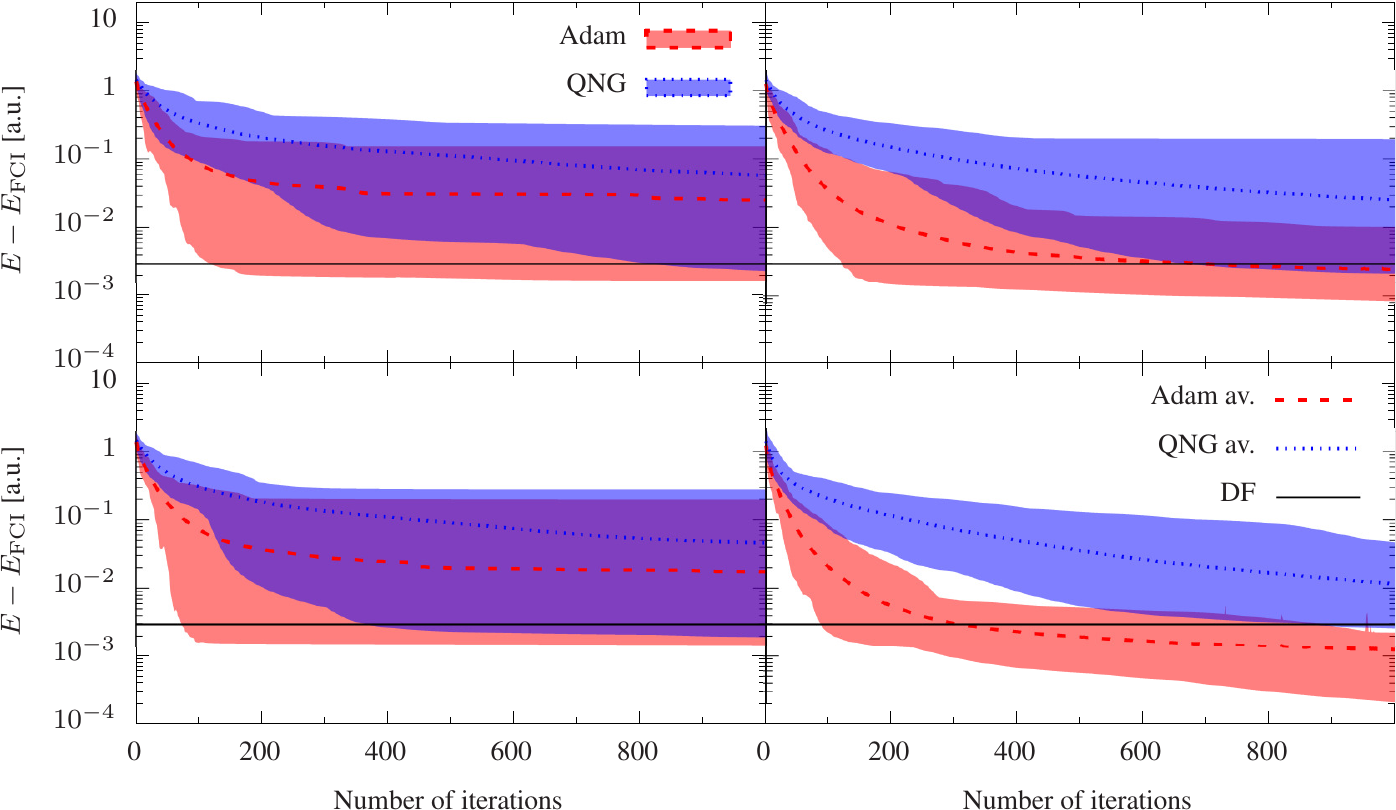}
\caption{
The difference between the expectation value~\eqref{eq_expval} obtained with the use of the HE ansatz with 5 layers and the result of the full configuration-interaction (FCI) calculation as a function of the number of iterations.
The calculations are performed for $A$ basis, consisting of 8 orbitals mapped to 6 qubits.
Single-qubit rotations $U^{(1q)}_{zx}$ and $U^{(1q)}_{zyz}$ are used for the first and second rows of panels, respectively.
The left panels correspond to the ansatz with merged entangling layers, while the right panels are related to the splitted ones.
Each panel displays the results of the set of calculations, whose size is equal to the total number of parameters in ansatz divided by 2.
In each calculation, the initial values of the parameters $\btheta$ are randomly chosen.
The values averaged over the sets of calculations are displayed with the dashed and dotted lines for the Adam and QNG optimizers, respectively.
The energy corresponding to the Dirac-Fock (DF) approximation is displayed with black solid line.
}
\label{fig_he_ansatz_5_layers}
\end{figure}

We start by investigating the performance of various HE ansatze and optimizers on the example of the smallest $A$ basis (see Table~\ref{tb_basis}).
The initial values of $\btheta$ parameters do not contain any physical meaning and are chosen arbitrary.
The calculations for each ansatz were run for several different initial values of $\btheta$.
The number of runs was chosen to be equal to the total number of parameters in the ansatz, divided by two.
For each run, the optimization procedure was interrupted after 1000 iterations.
Figure~\ref{fig_he_ansatz_5_layers} presents the dependence of the expectation value on the number of iterations for four different HE ansatze with 5 layers.
From the figure it is seen that the Adam optimizer converges faster than the QNG.
Moreover, the energy obtained with the Adam optimizer is closer to the exact value $E_{\rm FCI}$ than the one obtained with QNG.
From Fig.~\ref{fig_he_ansatz_5_layers} one can conclude that the ansatz with splitted entangling layers and single-qubit rotations $U_{zyz}^{(1q)}$ (bottom right panel) provides the best results.
\\
\indent
Let us study the dependence of the results on the number of layers $L$ in the ansatz.
Table~\ref{he_layers} presents the minimum difference $E_{\rm HE} - E_{\rm FCI}$ for $L$ 
layers in the HE ansatz obtained after a number of runs.
Each run was performed with different randomly chosen initial parameters $\btheta$ and consisted of 1000 optimization steps.
\begin{table}
\caption{
The minimum difference $E_{\rm HE} - E_{\rm FCI}$ in a.u. for $L$ layers in the HE ansatz obtained after a number of runs with randomly chosen initial parameters $\btheta$. For each run 1000 optimization steps were performed.
For each ansatz structure the number of runs equals to the total number of parameters divided by 2.
% The first ans second rows of cells correspond to the results obtained with the Adam and QNG optimizers, respectively.
}
\begin{ruledtabular}
\begin{tabular}{ccccc}
  & \multicolumn{2}{c}{merged} & \multicolumn{2}{c}{splitted}
\\
L & $U^{(1q)}_{zx}$ & $U^{(1q)}_{zyz}$ & $U^{(1q)}_{zx}$ & $U^{(1q)}_{zyz}$ 
\\ \hline
%   &      24 &      36 &      42 &      63 \\
%   &      30 &      45 &      54 &      81 \\
%   &      36 &      54 &      66 &      99 \\
%   &      42 &      63 &      78 &     117 \\
%
%  22 20.04 19:34
%  26 21.04 09:42 14.1 / 4
%  64 26.04 20:55 
% 232 16.05 08:15 
% 
% old data
% 
%   & \multicolumn{4}{c}{Adam} \\
% 3 & 2.5e-02 & 8.1e-03 & 3.9e-03 & 2.1e-03 \\
% 4 & 2.8e-02 & 2.1e-02 & 2.0e-03 & 1.7e-03 \\
% 5 & 2.6e-02 & 1.2e-02 & 2.4e-03 & 1.3e-03 \\
% 6 & 2.6e-02 & 2.2e-02 & 2.4e-03 & 1.3e-03 \\
%   & \multicolumn{4}{c}{QNG} \\
% 3 & 4.5e-02 & 2.3e-02 & 1.7e-02 & 1.1e-02 \\
% 4 & 6.3e-02 & 5.3e-02 & 1.4e-02 & 1.0e-02 \\
% 5 & 5.3e-02 & 4.0e-02 & 2.4e-02 & 1.0e-02 \\
% 6 & 7.8e-02 & 5.9e-02 & 2.7e-02 & 1.4e-02 \\
% 
% new data
% 
%  & \multicolumn{4}{c}{Adam} \\
%3 & 5.0e-03 & 4.3e-03 & 2.8e-03 & 2.1e-03 \\
%4 & 3.4e-02 & 1.2e-02 & 2.8e-03 & 1.5e-03 \\
%5 & 2.5e-02 & 1.8e-02 & 2.5e-03 & 1.2e-03 \\
%6 & 3.3e-02 & 1.5e-02 & 2.2e-03 & 1.2e-03 \\
%  & \multicolumn{4}{c}{QNG} \\
%3 & 3.3e-02 & 3.7e-02 & 1.8e-02 & 1.2e-02 \\
%4 & 4.6e-02 & 3.5e-02 & 1.6e-02 & 1.2e-02 \\
%5 & 5.8e-02 & 4.6e-02 & 2.5e-02 & 1.2e-02 \\
%6 & 9.5e-02 & 5.9e-02 & 2.4e-02 & 1.2e-02 \\
%
% best
& \multicolumn{4}{c}{Adam} \\
3  & 1.5[-3] & 1.6[-3] & 1.4[-3] & 3.8[-4] \\
4  & 1.5[-3] & 1.4[-3] & 1.1[-3] & 1.6[-4] \\
5  & 1.5[-3] & 1.7[-3] & 1.0[-3] & 4.6[-4] \\
6  & 1.7[-3] & 1.4[-3] & 7.0[-4] & 3.1[-4] \\
& \multicolumn{4}{c}{QNG} \\
3  & 1.8[-3] & 2.0[-3] & 2.6[-3] & 2.3[-3] \\
4  & 1.4[-3] & 2.5[-3] & 1.7[-3] & 2.4[-3] \\
5  & 2.1[-3] & 2.2[-3] & 2.3[-3] & 2.7[-3] \\
6  & 2.0[-3] & 2.0[-3] & 3.1[-3] & 2.0[-3] \\
\end{tabular}
\label{he_layers}
\end{ruledtabular}
\end{table}
From the table, it is seen that the ansatz with the splitted entangling layers and single-qubit rotations $U_{zyz}^{(1q)}$ allows one to obtain the closest to exact value result.
One can also observe that for all ansatze the best results are obtained with the Adam optimizer.
This can be explained by the faster convergence with respect to the number of iterations (see also Fig.~\ref{fig_he_ansatz_5_layers}) and, probably, by an ability of the Adam optimizer to escape some local minima.
The table shows that increasing the number of layers does not improve the result. 
%Indeed, the growth of the difference $\bar{E}_{\rm HE} - E_{\rm FCI}$ with the increase of $L$ indicates the increase of the number of bad initial parameters.
%This, in turn, leads to the growth of the number of runs needed to achieve good results for a given ansatz.
Moreover, the number of runs needed to achieve the result with the same accuracy grows with $L$.
Therefore, we did not perform calculations with the HE ansatz for the basis sets $B$ and $C$, which require more qubits and, consequently, a larger number of parameters $\btheta$.
It is also worth mentioning that due to the overall bad accuracy obtained with the HE ansatz, we don't discuss the problem of the vanishing gradients, the so-called barren plateau issue, which drastically slows down or even eliminates the convergence. 
The origins of this problem as well as the possible ways to overcome it can be found, e.g., in Refs.~\cite{McClean_NC9_4812_2018, Grant_Wossnig_, Verdon_, Cerezo_Sone_, Arrasmith_Cerezo_, Rad_Seif_Linke}.
%
% - - - - - - - - - - - - - - - - - - - - - - - - - - - - - - - - - -
%
\subsubsection{dUCC-SD}
\begin{figure}
\centering
\includegraphics[width=\textwidth]{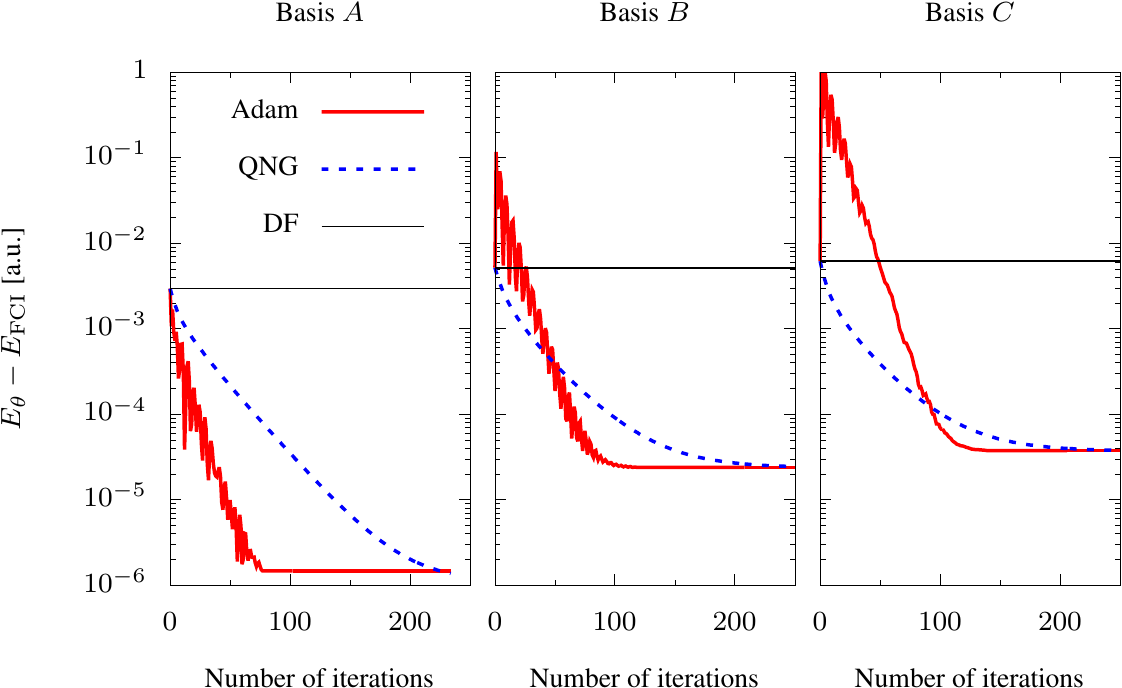}
\caption{
The difference between the expectation value~\eqref{eq_expval} obtained with the use of the dUCC-SD ansatz and the result of the full configuration-interaction (FCI) calculation as a function of the number of iterations. 
The energy corresponding to the Dirac-Fock (DF) approximation is displayed with black solid line.
}
\label{fig_ducc_ansatz}
\end{figure}

Let us turn to the consideration of the dUCC-SD ansatz.
For this ansatz the Dirac-Fock wave function of the state $\left(7s^2 7p^2_{1/2} 7p_{3/2}\right)_{m = 1/2}$ was used as the initial wave function $\ket{\Psi_0}$.
The dUCC ansatz conserves the projection of the total angular momentum $m$, and thus the energy minimization is performed in the space of the $\hat{J}^2$ operator's eigenstates with the different total angular momentum $J$ but the same $J_z = m$.
Therefore, one needs to set $m$ equal to the minimal allowed projection for the electronic problem under consideration. %{\rd Why?}
%The initial values of the parameters $\btheta = 0$ that allows to start the search in a region of the parameter space that is close to the optimum, provided by the Dirac-Fock solution.
%
The initial values of the parameters~$\theta$ were set to zero that allowed us to start searching near 
the Dirac-Fock solution~$\ket{\Phi_0}$. 
\\
\indent
Figure~\ref{fig_ducc_ansatz} presents the difference $E_{\btheta} - E_{\rm FCI}$ as a function of the number of iterations for basis sets $A$, $B$, and $C$.
From this figure, it is seen that for all basis sets it takes about 100 iterations for the Adam optimizer to converge. 
The QNG optimizer, while showing monotonic convergence in the logarithmic scale, requires more iterations than the Adam one.
\\
\indent
Table~\ref{ducc_data} presents various parameters and results of the calculations performed with the dUCC-SD ansatz.
\begin{table}
\caption{
The number of parameters corresponding to the single $N_{\btheta}^{(S)}$ and double $N_{\btheta}^{(D)}$ excitations for $A$, $B$, and $C$ basis sets (see Table~\ref{tb_basis}).
The size of the configuration space, i.e., the number of the Slater determinants with the odd parity and angular momentum projection $m = 1/2$, generated by the dUCC-SD ansatz $N_{\rm Sl.\ det.}^{\rm (SD)}$ is presented in the fourth column.
% {\rd I'm not sure that this is correct} 
In the fifth column, the total size of the configuration space $N_{\rm Sl.\ det.}^{\rm (FCI)}$ for given quantum numbers is presented.
}
\begin{ruledtabular}
\begin{tabular}{crrrrc}
Basis &  $N_{\btheta}^{(S)}$ & $N_{\btheta}^{(D)}$ & $N_{\rm Sl.\ det.}^{\rm (SD)}$ & $N_{\rm Sl.\ det.}^{\rm (FCI)}$ & $E_{\btheta} - E_{\rm FCI}$ [a.u.] \\
\hline
  $A$ &  1 &   4 &       7 &       7 & $1.4 \times 10^{-6}$ \\
  $B$ &  9 &  68 &     450 &     472 & $2.4 \times 10^{-5}$ \\
  $C$ & 13 & 565 &  485820 &  500018 & $3.8 \times 10^{-5}$ \\
\end{tabular}
\label{ducc_data}
\end{ruledtabular}
\end{table}
From this table it is seen that the difference $E_{\btheta} - E_{\rm FCI}$ for all basis sets is several orders of magnitude more precise than the one obtained with the HE ansatz for the smallest $A$ basis.
It is also worth to stress that, in contrast to the HE ansatz, in the dUCC ansatz the initial values of the parameters $\btheta = 0$ and, as a result, multiple runs are not required.
% The table demonstrates the well-known fact that the exponential growth of the configuration space with the increase of the system size can be effectively described in the framework of the coupled cluster approach by the SD excitations.
%The number of parameters $\btheta$ meanwhile grows only polynomially, and the corresponding accuracy of the energy obtained in the dUCC-SD approach almost does not decrease.
The number of parameters $\btheta$ meanwhile grows only polynomially, 
and the difference between FCI and dUCC-SD values almost does not increase.
It is also important to emphasize that the number of iterations does not change significantly, despite a two orders of magnitude increase in the number of varying parameters in the basis $C$ with respect to the basis $A$.
Moreover, the growth of the quantum resorces required for the implementation of the dUCC-SD ansatz scales only polynomially with the number of qubits.
%Additionally, we note that the dUCC ansatz can be effectively applied on a quantum computer, which turns out to be impossible in the case of a classical one.
%With all these in mind we expect that the simulation of the electronic structure of various ions and atoms on upcoming quantum devices by the VQE algorithm with the dUCC ansatz will allow one to obtain the results for the systems unreachable for classical computers.
Therefore one can expect that the VQE algorithm with the dUCC ansatz may allow one to obtain results for the configuration
space unreachable for the classical computers, provided that 
the execution of the required circuits on the quantum device is possible.   
%
% -------------------------------------------------------------------
%
\subsection{iPEA}
\begin{figure}
\centering
\includegraphics[width=\textwidth]{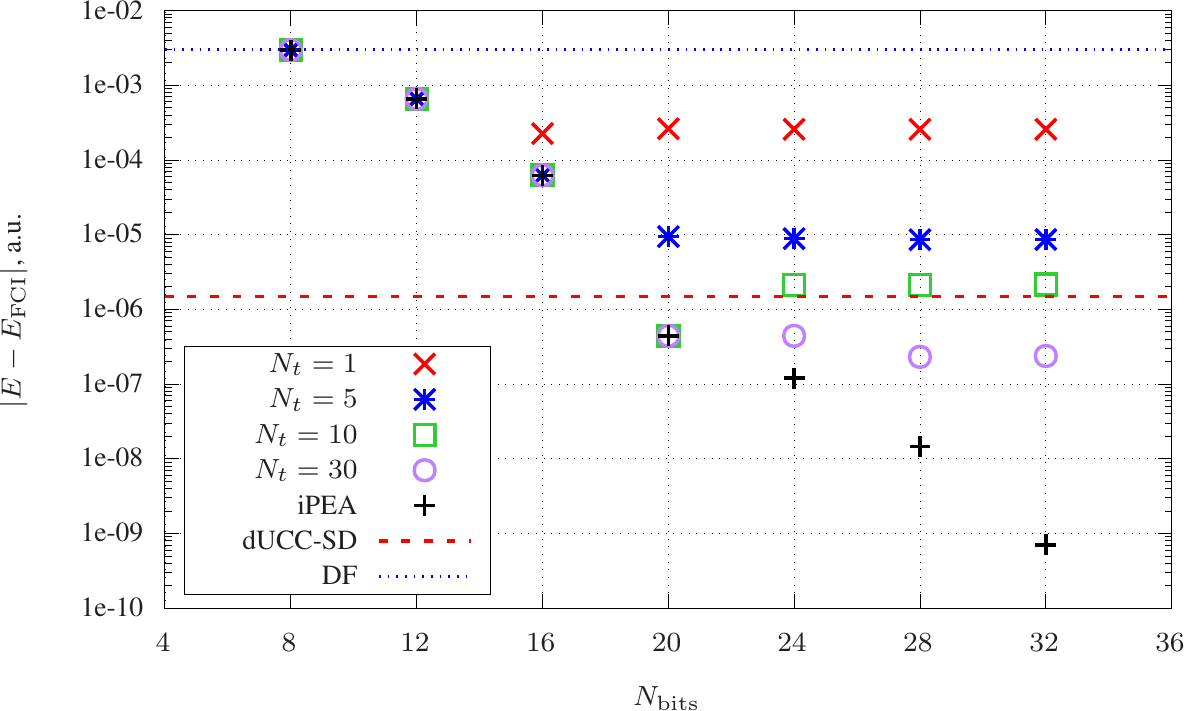}
\caption{
The difference between the exact ground-state energy $E_{\rm FCI}$ and the energy obtained by iPEA as a function of the number of measured bits $N_{\rm bits}$.
The results of the iPEA without the Trotterization are represented by black plus symbols.
The calculations were performed for the $A$ basis, $\Delta E = 3\pi$ a.u., and $E' = E_{\rm DF}$.
The accuracy obtained by the VQE algorithm with dUCC-SD ansatz is indicated with the red dashed line.
}
\label{fig_ipea_trotter}
\end{figure}

We now turn to the evaluation of the ground-state energy of the moscovium atom by the iPEA.
Let us investigate the dependence of the Trotterization error on $N_t$ for the smallest $A$ basis.
For this purpose, in Fig.~\ref{fig_ipea_trotter} we present the difference between the exact ground-state energy and the energy obtained by the iPEA with various $N_t$.
Here we used the Dirac-Fock wave function as $\ket{\Psi'}$, set $E' = E_{\rm DF}$, and fix $\Delta E = 3\pi$ a.u. to guarantee $t < 1$ [see Eq.~\eqref{eq_ipea_t}] that allows us to assume the applicability of the approximation~\eqref{eq_evolution}.
From this figure one can see that the Trotter error can be expressed in the maximal number of bits which are measurable
for given $N_t$.
Indeed, the use of the approximation~\eqref{eq_evolution} corresponds to the change of the Hamiltonian $\hat{H}$ with $\hat{H}'$ defining by Eq.~\eqref{eq_evolution_trot}.
As a result, the iPEA does not allow us to measure the exact eigenvalue of $H$, instead its approximation provided by 
the corresponding eigenvalue of the Hamiltonian $\hat{H}'$ is measured.
In particular, one can find the ground-state energy as the eigenvalue of the Hamiltonian $\hat{H}'$ 
corresponding to its eigenstate with the largest overlap with the initial state $\ket{\Psi'}$.
The difference of this eigenvalue with the ground-state energy is large for small $N_t$ and small number of bits is worth measurement.
With the growth of $N_t$ the accuracy increases and one can determine more digits of the ground-state energy.
From Fig.~\ref{fig_ipea_trotter} it is seen that with $N_t = 1$, one can only slightly improve the accuracy of the Dirac-Fock approximation.
The accuracy of the VQE algorithm with dUCC-SD ansatz can be obtained with $N_t \geqslant 10$.
We note that instead of increasing $N_t$ one can utilize the Suzuki-Trotter formulas of the higher orders, thus reducing the Trotter error.
The study of this approach, however, lies beyond the scope of the present investigation.
%
% ===================================================================
%
% \subsection{Estimation of the number of gates}
% \subsection{Comparison of VQE and iPEA}
\subsection{Number of gates required for the VQE and iPEA}
Let us compare the number of gates required for the ground-state energy calculation by VQE and iPEA. 
For this purpose, in Table~\ref{tb_gates} we present the number of one- and two-qubit gates used in these two algorithms in the case of the smallest $A$ basis.
\begin{table}
\caption{
The number of the single-qubit (SQ) and CNOT gates required for the calculation of the moscovium ground-state energy in the $A$ basis.
In the third row, the data for the splitted HE ansatz with number of layers $L=5$ and single-qubit rotations $U^{(1q)}_{zyz}$
are presented.
In the forth row, the number of gates for dUCC-SD ansatz transformed into the qubit operations by the conventional ``greedy'' approach and within the strategy from Refs.~\cite{Cowtan, Vandaele, Patel} are given.
The data for the trotterized evolution $e^{i\hat{H}'\tau}$ and controlled trotterized evolution $Ce^{i\hat{H}'\tau}$ are presented in last two rows.
To estimate the number of gates for $Ce^{i\hat{H}'\tau}$, each operation which was used in the trotterized evolution is converted into controlled gates by the relations presented in Ref.~\cite{Barenco}.
}
\begin{ruledtabular}
\begin{tabular}{ccccc}
 &  \multicolumn{2}{c}{``greedy''} &  \multicolumn{2}{c}{strategy from Refs.~\cite{Cowtan, Vandaele, Patel}} \\
 &  SQ & CNOT &  SQ & CNOT\\
\hline
HE                   & 198 &  25 &     &     \\
dUCC-SD              & 198 & 156 &  76 &  88 \\
$e^{i\hat{H}'\tau}$  & 604 & 568 & 192 & 237  \\
% Naive: Rz = 124, V = 232, H = 248, cnot = 568
% CVP: Rz = 124, V = 32, H = 36, cnot = 237
$Ce^{i\hat{H}'\tau}$ & 6552 & 4368 & 2549 & 1770  \\
% CRz = 2 cnot + 2 SqC
% CH = 1 cnot + 2 SqC
% CV = 2 cnot + 3 SqC
% Ccnot = 6 cnot + 9 SqC
\end{tabular}
\label{tb_gates}
\end{ruledtabular}
\end{table}
From the table one can see that the dUCC-SD ansatz, compiled in accordance with the strategy from Refs.~\cite{Cowtan, Vandaele, Patel}, requires less one-qubit but more two-qubit CNOT gates when compared to the HE ansatz.
Here we have assumed all-to-all connectivity of the qubits, which is not feasible in many quantum computers architectures.
Account for the absence of such connectivity results in the growth of the number of two-qubit gates in the dUCC-SD ansatz.
The circuit for the dUCC-SD ansatz is, therefore, deeper than one of the HE ansatz and more prone to decoherence and other types of noise.
In course of the non-fault tolerant calculations a part of such noise can be suppressed with the use of error mitigation techniques~\cite{Temme_PRL119_180509_2017, Nation, Li_Benjamin, Kandala_N567_491_2019, Carbone, Yeter_Aydeniz}.
However, the realization of the dUCC-SD ansatz for the studied system on present-day quantum computers is unfeasible even with these techniques.
From the other side, the HE ansatz requires multiple runs with random initial parameters and much larger number of the optimization steps.
Moreover, with this ansatz we failed to achieve sufficiently accurate result even in the case of the smallest $A$ basis.
We conclude that the HE ansatz, though requiring shallow circuits with relatively small number of gates, is inappropriate for the calculation of the ground-state energies of atomic systems.
\\
\indent
In Table~\ref{tb_gates} the number of gates for the controlled Trotterized evolution $Ce^{i\hat{H}'\tau}$, which is required for the iPEA, is also presented.
Here it is worth to mention that in order to apply the compilation strategy from Refs.~\cite{Cowtan, Vandaele, Patel} it is necessary to partition the Hamiltonian $\hat{H}$ into the sets of mutually commuting Pauli strings before the Trotterization~\eqref{eq_evolution}.
Partitioning in the minimal number of such sets is the NP-hard problem~\cite{np_hard}, that strongly limits the application of the compilation strategy for the Hamiltonians of systems intractable by the conventional (classical) computers.
Measurement of $N_{\rm bits}$ bits of the energy by the iPEA requires application of $N_t 2^{N_{\rm bits}}$ controlled Trotterized evolutions.
In the case of the $A$ basis, one needs to use $N_t \sim 10$ and $N_{\rm bits} = 20$ in order to obtain the accuracy comparable to the VQE with the dUCC-SD ansatz.
At these parameters, the number of gates required for the iPEA is several orders of magnitude larger than the number of gates needed for the VQE algorithm.
It leads to the conclusion that the VQE algorithm with the dUCC-SD ansatz is the most perspective for the calculation of the ground-state energies of the atomic systems.
%
% ===================================================================
%
\section{Conclusion}
The possibility to simulate the electronic structure of ions and atoms by quantum algorithms has been studied on the example of the moscovium atom.
The ground-state energy of the moscovium was calculated with the use of the VQE and iPEA. 
\\ 
\indent
Computations by the VQE were performed with problem inspired dUCC-SD and hardware efficient HE ansatze.
In the case of the HE ansatz, we have investigated the dependence of the accuracy of the calculated energy on the structure of the ansatz, number of layers, and types of the one-qubit rotations.
It was found that this ansatz doesn't allow to significantly improve the accuracy of the Dirac-Fock approximation.
We note, that the calculations with HE ansatz are also hindered by the necessity in multiple runs with different arbitrary initial parameters.
The dUCC-SD ansatz, in contrast, possesses well-defined initial values of the parameters and allows one to obtain energy with the high precision in the relatively small number optimization steps.
Moreover, we have found that the calculations with dUCC-SD ansatz remain stable as the numbers of basis functions and active electrons increase.
The accuracy of the energy obtained with this ansatz for the basis sets corresponding to the configuration spaces with about 10 and 500000 Slater determinants was found to be comparable and achievable in about 100 optimization steps.
\\
\indent
We have also studied the performance of the Adam and QNG optimizers within the framework of the VQE algorithm.
It was found that in the case of the HE ansatz the Adam optimizer converges to more accurate values of the energy than the QNG does. 
In the case of the dUCC-SD, both optimizers provide the results on the same level of accuracy.
We note that for this ansatz the QNG demonstrates uniform convergence and, for the largest basis set, requires around 100 optimization steps as the Adam does.
This reflects the good scalability of the QNG with the number of parameters.
However, in addition to the gradients, QNG requires the measurement of the Hessian on the quantum computer that makes this optimizer much more expensive in terms of the quantum computer resources than the Adam.
\\
\indent
We have also evaluated the ground-state energy of the moscovium by iPEA.
For this purpose we have utilized the first-order Suzuki-Trotter formula to transform the evolution operator into the form suitable for the conversion into the gates.
The dependence on the Trotter number $N_t$, which defines the precision of the Trotterization, was studied.
It was found that in order to achieve the accuracy of the VQE algorithm with dUCC-SD ansatz by the iPEA one needs to use $N_t \sim 10$.
At such $N_t$ the number of required gates is several orders of magnitude larger than the number of gates required for the construction of the dUCC-SD ansatz.
\\
\indent
With all this in mind, one can conclude that the VQE algorithm with the dUCC-SD ansatz and Adam optimizer is the most promising 
approach for the calculation of the ground-state energies of the various atomic systems on the near-term quantum computers.
%
% ===================================================================
%
\section*{Acknowledgements}
We thank  Yu.~Ts.~Oganessian for drawing our attention to the problem considered in the paper.
We also thank M.~Y.~Kaygorodov for providing us with the matrix elements in the Dirac-Fock basis and E.~Eliav, V.~V.~Korenkov, and S.~V.~Ulyanov for fruitful discussions.
The work is supported by the Ministry of Science and Higher Education of the Russian Federation within the Grant No. 075-10-2020-117.
%
% ===================================================================
%
\appendix
\section{Transformation of the commuting Pauli exponentials to the gates}
\label{appendix}
Let us briefly describe the conventional ``greedy'' approach and the strategy from Refs.~\cite{Cowtan, Vandaele, Patel} for the conversion of the Pauli exponentials
\begin{equation}
e^{-i\sum_n \alpha_n P_n / 2} = \Pi_n e^{-i \alpha_n P_n / 2}
\label{eq_p_exp}
\end{equation}
into the gates.
Here it is assumed that all Pauli strings $P_n$ commute with each other.
Such Pauli exponentials naturally appear in the dUCC ansatz and can be formed when applying the Trotterization to the evolution operator.
\\
\indent
Within the so-called ``greedy'' approach~\cite{Nielsen_Chuang}, each Pauli exponential is transformed into the gates separately.
For the sake of clarity, let us describe this transformation on the example of
\begin{equation}
P_n = \sigma^x_i \otimes \sigma^y_j \otimes \sigma^z_k.
\end{equation}
First, one diagonalizes each $\sigma^x$ and $\sigma^y$ which appears in $P_n$ by conjugating the Pauli exponential
\begin{equation}
e^{-i \alpha_n P_n/2} = G e^{-i \alpha_n P^{(z)}_n/2} G^\dagger.
\end{equation}
Here $P^{(z)}_n$ designates $P_n$ in which $\sigma^x$ and $\sigma^y$ are replaced with $\sigma^z$ and $G \in \lbrace I,H,V\rbrace^{\otimes N_q}$ stands for the tensor product of the single-qubit Cliffords with
\begin{equation}
V = \frac{1}{\sqrt{2}}\begin{pmatrix}1 &-i \\-i & 1 \end{pmatrix}.
\end{equation}
In our case
\begin{equation}
G = H_i\otimes V_j,
\qquad
P^{(z)}_n = \sigma^z_i \otimes \sigma^z_j \otimes \sigma^z_k.
\end{equation}
The diagonalized Pauli exponential acts as a phase gate on a qubit state
\begin{equation}
e^{-i\alpha_nP^{(z)}_n/2}\ket{\bq} = e^{-i\alpha_n\lambda^{(\bq)}_n/2}\ket{\bq},
\end{equation}
where $\lambda_n^{(\bq)}$ stands for the parity of the qubits on which $P^{(z)}_n$ acts on with $\sigma^z$.
Therefore, one can apply diagonalized Pauli exponential by collecting the parity to one of the qubits with the CNOT gates
\begin{equation}
{\rm CNOT}\ket{x}\ket{y} = \ket{x}\ket{x\oplus y},
\end{equation}
applying $R_z$ gate to this qubit, and restoring the parity of the qubits.
In our example, $\lambda_n^{(\bq)} = q_i \oplus q_j \oplus q_k$ and the diagonalized Pauli exponential is given by the following quantum circuit
\begin{center}
\includegraphics[width=0.4\textwidth]{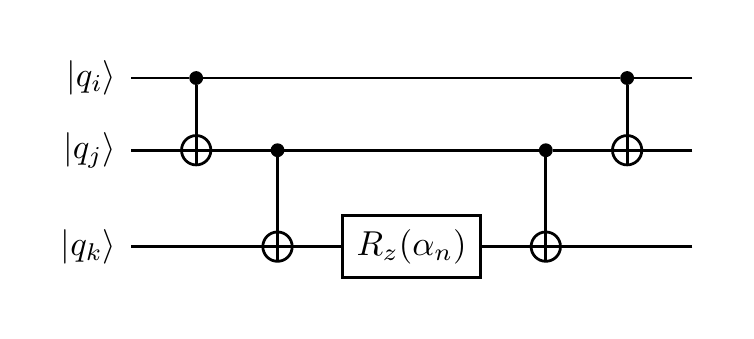}
\end{center}
This completes the ``greedy'' strategy for the transformation of the Pauli exponentials into gates.
\\
\indent
Alternatively, one can utilize the three-step strategy from Refs.~\cite{Cowtan, Vandaele, Patel}, which reduces the number of gates required in the ``greedy'' approach.
The fundamental difference between this strategy and the conventional approach is that all commuting Pauli exponentials are processed simultaneously. 
In the first step of the strategy, all PSs are simultaneously diagonalized 
\begin{equation}
e^{-i\sum_n \alpha_n P_n/2}
=
G
e^{-i\sum_m \beta_m P^{(z)}_m/2}
G^\dagger.
\label{eq_exp_p}
\end{equation}
Here $G$ is the unitary operation constructed in accordance with the algorithm from Ref.~\cite{Cowtan} and consisting of CNOT gates and single-qubit Cliffords $H$ and $V$.
Note that $P^{(z)}_m$ may differ from ones obtained by a simple replacement in $P_n$ of $\sigma^x$ and $\sigma^y$ with $\sigma^z$.
In the second step, in accordance with the algorithm from Ref.~\cite{Vandaele}, one constructs the quantum circuit consisting of CNOT and $R_z$ gates, which acts on the qubits' state as follows
\begin{equation}
e^{-i\sum_m \beta_m P^{(z)}_m/2}
\ket{\bq}
= 
e^{-i\sum_m \beta_m \lambda^{({\bq})}_m} 
\ket{A\bq},
\end{equation}
where $A$ is the linear reversible $N_q \times N_q$ matrix with elements in the two-element field $\mathbb{F}_2 = \lbrace0,1\rbrace$.
From this expression it is seen that the constructed quantum circuit provides the correct phase factor but modifies the qubits' state.
To restore the qubits' states, in the third step, one applies $A^{-1}$ consisting exclusively from CNOT gates as described in Ref.~\cite{Patel}.
%
% ======================================================================
%
\section{Adam and QNG optimizers}
\label{appendix_adam_qng}
The Adam optimizer~\cite{Kingma_Ba} is the first-order gradient descent method, in which the parameters $\btheta$ are updated in accordance with the rule
\begin{equation}
\theta^{(n+1)}_i = \theta^{(n)}_i - \frac{\eta}{\sqrt{v^{(n)}_i} + \epsilon} m_i^{(n)},
\label{eq_adam}
\end{equation}
where $n$ stands for the iteration step, $\eta$ is the learning rate, $\epsilon$ is the regularization constant, and the exponentially moving averages of the gradient $m_i^{(n)}$ and of the squared gradient $v^{(n)}_i$ are defined as
\begin{equation}
m^{(n)}_i 
= 
\frac{\beta_1 - \beta_1^n}{1 - \beta_1^n}m^{(n-1)}_i
+
\frac{1 - \beta_1}{1 - \beta_1^n}\frac{\partial E(\btheta^{(n)})}{\partial \theta^{(n)}_i},
\end{equation}
\begin{equation}
v^{(n)}_i 
= 
\frac{\beta_2 - \beta_2^n}{1 - \beta_2^n}v^{(n-1)}_i
+
\frac{1 - \beta_2}{1 - \beta_2^n}
\left[\frac{\partial E(\btheta^{(n)})}{\partial \theta^{(n)}_i}\right]^2,
\end{equation}
with $m^{(0)} = 0$, $v^{(0)} = 0$, and $\beta_{1,2}$ designating the hyperparameters. 
In the present investigation we fix $\eta = 0.05$, $\epsilon = 10^{-8}$, $\beta_1 = 0.9$, and $\beta_2 = 0.999$.
\\ \indent
In the QNG optimizer, the updated values of the parameters $\btheta^{(t+1)}$ are defined as the solution to the equation~\cite{Stokes_Q4_269, Wierichs_PRR2_043246_2020}
\begin{equation}
F^{(t)}
\left(\btheta^{(t+1)} - \btheta^{(t)}\right) = -\eta \bnabla E(\btheta^{(t)}),
\label{eq_natgrad}
\end{equation}
with
\begin{equation}
F^{(t)}_{ij} = 
\Re\left[\braket{\frac{ \partial \Psi(\btheta^{(t)})}{\partial \theta^{(t)}_i}}{\frac{\partial \Psi(\btheta^{(t)})}{\partial \theta^{(t)}_j}}\right]
-
\braket{\frac{\partial \Psi(\btheta^{(t)})}{\partial \theta^{(t)}_i}}{\Psi(\btheta^{(t)})}
\braket{\Psi(\btheta^{(t)})}{\frac{\partial \Psi(\btheta^{(t)})}{\partial \theta^{(t)}_j}}
\end{equation}
standing for Fubini-Study metric tensor which reflects the geometry of the quantum states.
Most often then not, equation~\eqref{eq_natgrad} is undetermined and $F^{(t)}$ cannot be inverted.
To solve this ill-posed inverse problem we apply Tikhonov regularization and search for such $\btheta^{(t+1)}$ that minimizes the functional
\begin{equation}
\mathcal{F}(\lambda) = \left\Vert
F^{(t)}
\left(\btheta^{(t+1)} - \btheta^{(t)}\right) + \eta \bnabla E(\btheta^{(t)})
\right\Vert^2 
+ 
\lambda \left\Vert \btheta^{(t+1)} - \btheta^{(t)} \right\Vert^2.
\end{equation}
The presence of the regularization parameter $\lambda$ guarantees the smooth variation of the parameters $\btheta$.
At each time step the value of $\lambda$ is selected as the L-curve corner and calculated in accordance with the algorithm suggested in Ref.~\cite{Cultrera_Callegaro}.
Note that in the limit $\eta \rightarrow 0$ the update of $\btheta$ by the QNG optimizer corresponds to the imaginary time evolution in the variational space~\cite{McArdle_npjQI5_75_2019}.
Here, as for the Adam optimizer, we fix $\eta = 0.05$.
%
% ===================================================================
%

\end{document}